\documentclass[prb,twocolumn,superscriptaddress,preprintnumbers,a4paper,amsmath,amssymb,showpacs,floatfix]{revtex4}

\usepackage{graphicx} 
\usepackage{bm}

\begin{document}
 
\title{Transition from a phase-segregated state to single-phase incommensurate sodium ordering in $\gamma$-Na$_x$CoO$_2$ with $x \approx 0.53$} 
\author{R.~Feyerherm}
\email[Email of corresponding author: ]{ralf.feyerherm@helmholtz-berlin.de} 
\affiliation{Helmholtz-Zentrum Berlin GmbH f\"ur Materialien und Energie, BESSY, 12489 Berlin, Germany}
\author{E.~Dudzik} 
\affiliation{Helmholtz-Zentrum Berlin GmbH f\"ur Materialien und Energie, BESSY, 12489 Berlin, Germany}
\author{S.~Valencia} 
\affiliation{Helmholtz-Zentrum Berlin GmbH f\"ur Materialien und Energie, BESSY, 12489 Berlin, Germany}
\author{A.~U.~B.~Wolter}
\altaffiliation{now at IFW, 01069 Dresden, Germany}
\affiliation{Helmholtz-Zentrum Berlin GmbH f\"ur Materialien und Energie, BESSY, 12489 Berlin, Germany}
\affiliation{Institut f\"{u}r Physik der Kondensierten Materie, Technical University Braunschweig, 38106 Braunschweig, Germany}
\author{C.~J.~Milne} 
\affiliation{Helmholtz-Zentrum Berlin GmbH f\"ur Materialien und Energie, 14109 Berlin, Germany} 
\author{S.~Landsgesell} 
\affiliation{Helmholtz-Zentrum Berlin GmbH f\"ur Materialien und Energie, 14109 Berlin, Germany}
\author{D.~Alber} 
\affiliation{Helmholtz-Zentrum Berlin GmbH f\"ur Materialien und Energie, 14109 Berlin, Germany}
\author{D.~N.~Argyriou} 
\affiliation{Helmholtz-Zentrum Berlin GmbH f\"ur Materialien und Energie, 14109 Berlin, Germany}

\preprint{}

\date{\today}


\begin{abstract}
Synchrotron X-ray diffraction investigations of two single crystals of sodium cobaltate Na$_{x}$CoO$_2$ from different batches with composition $x = 0.525 - 0.530$ reveal homogeneous incommensurate sodium ordering with propagation vector (0.53~0.53~0) at room-temperature. The incommensurate $(q q 0)$ superstructure exists between 220~K and 430~K. The value of $q$ varies between $q = 0.514$ and 0.529, showing a broad plateau at the latter value between 260~K and 360~K. On cooling, unusual reversible phase-segregation into two volume fractions is observed. Below 220~K, one volume fraction shows the well-known commensurate orthorhombic $x = 0.50$ superstructure, while a second volume fraction with $x = 0.55$ exhibits another commensurate superstructure, presumably with a $6a \times 6a \times c$ hexagonal supercell. We argue that the commensurate-to-incommensurate transition is an intrinsic feature of samples with Na concentrations $x = 0.5 + \delta$ with $\delta \approx 0.03$.
\end{abstract}

\maketitle

\section{Introduction}

The alkali cobaltate $\gamma$-Na$_x$CoO$_2$ has raised considerable interest as model compound for the doping of triangular transition metal oxide layers in comparison with the well-known square layered cuprate high-temperature superconductors. Extensive research efforts were initiated by the observation of superconductivity with $T_c$ of up to 5~K in hydrated Na$_{0.35}$CoO$_2$.\cite{Takada:2003} In Na$_x$CoO$_2$ the basic effect of a variation of the sodium concentration  $x$ is a simultaneous control of the magnetic and electronic degrees of freedom of the quasi two-dimensional triangular CoO$_2$ sheets. Consequently, the system Na$_x$CoO$_2$ exhibits a rich phase diagram as function of $x$ with a wealth of interesting electronic and magnetic ground states.\cite{Foo:2004, Yokoi:2005, Mendels:2005} Metallic like resistivity down to low temperatures has been observed for a wide range of sodium concentrations, except for Na$_{0.5}$CoO$_2$, for which a metal-insulator transition is observed at $T_{c2} = 53$~K.

In addition to providing a way of doping the CoO$_2$ sheets, variation of the Na content has been shown to result in various types of ordered Na superstructures, as demonstrated first by electron diffraction studies on a broad series of Na$_x$CoO$_2$ samples with $x$ ranging from 0.15 to 0.75.\cite{Zandbergen:2004} 
Obviously, sodium ion ordering may affect the electronic and magnetic properties of these compounds considerably, because it imposes electronic constraints on the CoO$_2$ layer. For Na$_{0.5}$CoO$_2$, e.g., a very stable highly ordered orthorhombic crystal structure has been determined by neutron diffraction \cite{Huang:2004} in which two distinct Co sites are occupied in a 1:1 ratio. It has been argued that in this case the ordered Na potential may be the driving force for charge ordering.\cite{Zhang:2005} 

Due to its important role in determining the physical properties of the Na$_x$CoO$_2$ series, sodium superstructure formation has been subject to extensive theoretical and experimental investigations. The screened Coulomb interaction among Na ions was identified as primary driving force\cite{Zhang:2005} while the formation of rows of Na ions perpendicular to the $a$ axis with Na-Na distance $\sqrt 3 a$ has been proposed as guiding principle for Na ordering.\cite{Zandbergen:2004} The first $x$-ray diffraction study was carried out on a single crystal with $x = 0.75$ using high-energy synchrotron radiation. The observed superstructure was discussed in terms of charge-stripe formation.\cite{Geck:2006} Single crystal neutron diffraction studies of Na ordering on samples with nominal composition $x$ = 0.78, 0.85, and 0.92 revealed complex superstructures which were explained by a theoretical model that results in di-, tri-, and quadrivacancy cluster formation as the general construction scheme for $x \geq 0.75$.\cite{Roger:2007}

Most interestingly, in many studies $T$-dependent structural transitions related to sodium ion re-arrangement have been observed. An early neutron powder diffraction report\cite{Huang:2004a} on a Na re-arrangement transition around $T = 330$~K in samples with $x = 0.75$ may be related to a melting of a specific superstructure observed by high-energy x-ray diffraction.\cite{Geck:2006} In various samples with $x \approx 0.8$ (partial) melting of sodium superstructures or more complex transitions between ordered phases with different periodicities have been observed between 270~K and 290~K.\cite{Weller:2009, Foury:2009, Morris:2009, Huang:2009a} Several studies point to a high Na mobility above about 200 K.\cite{Weller:2009, Julien:2008, Lang:2008}

Although the stoichiometry $x = 0.5$ is regarded as forming a very stable orthorhombic superstructure,\cite{Huang:2004} several studies reported $T$-dependent structural transitions in various $x \approx 0.5$ samples. In the original electron diffraction work a Na$_{0.5}$CoO$_2$ sample developed an incommensurate (IC) modulation along the [110] direction after excessive electron irradiation. In addition, on cooling to about 100~K, a few areas of the sample showed extra reflections from a tripling of the orthorhombic unit cell.\cite{Zandbergen:2004} Related behavior was observed in a combined electron diffraction and Raman spectroscopy study, which revealed a series of temperature dependent structural transitions at $\approx 200$~K, 410~K, and 470~K in Na$_{0.5}$CoO$_2$.\cite{Yang:2005} Especially the observation of a transition from the orthorhombic low-temperature structure to an incommensurate phase around 200~K was surprising since no anomaly in this temperature range had been reported before for Na$_{0.5}$CoO$_2$. Incommensurate ordering has also been reported for a series of polycrystalline samples.\cite{Lang:2008, Platova:2009}
Therefore, a more thorough study of the proposed incommensurate phase appeared necessary. 

In this work, we report x-ray diffraction investigations of the sodium ordering in two single crystals of Na$_{x}$CoO$_2$ from different batches with nominal composition $x$ close to 0.5 for variable temperatures 10~K - 450~K. From the $c$-axis parameters we derive values of $x = 0.525 - 0.530$ for these samples. In both samples, an incommensurate superstructure with propagation vector (0.53~0.53~0) was observed at room-temperature, either after heat treatment (sample~\# 1) or immediately after synthesis (sample~\# 2).

On cooling, we observe unusual reversible phase-segregation into two volume fractions. Below 220 K, we find one volume fraction with the well-known\cite{Huang:2004} commensurate orthorhombic 
superstructure reported for $x = 0.50$, besides a second volume fraction with $x = 0.55$ that exhibits another commensurate superstructure, presumably with a $6a \times 6a \times c$ hexagonal supercell. The incommensurate $(q q 0)$ superstructure exists between 220~K and 430~K. The value of $q$ exhibits a temperature dependent variation between $q = 0.514$ and 0.529, showing a broad plateau at the latter value between 260~K and 360~K. We argue that the commensurate-to-incommensurate transition is an intrinsic feature of samples with Na concentrations $x = 0.5 + \delta$ with $\delta \approx 0.03$.

\section{Experimental}

A single crystal sample was obtained using the floating zone method to first synthesize a Na$_{0.75}$CoO$_2$ boule.  
From the boule single crystal pieces were selected and immersed in a 0.1 molar solution of Br$_2$ in acetonitrile with a isomolar ratio of Br to Na and a reaction time of two weeks at room temperature. The Na/Co ratio of the product was measured using neutron activation analysis (NAA). Two different batches gave nominal compositions of Na$_{0.48}$CoO$2$ (sample~\#1) and Na$_{0.42}$CoO$_2$ (sample~\#2). Resistivity measurements on a piece of crystal \#1 confirmed the presence of the charge localization transition at 52~K. 

For the x-ray diffraction experiments pieces of dimensions $0.15 \times 3 \times 6$~mm$^3$ were cleaved off, where the smallest dimension was parallel to the crystal $c$ axis. Synchrotron x-ray diffraction measurements were conducted on the MAGS beamline at HZB.\cite{Dudzik:2006} The experimental endstation consists of a six-circle Huber diffractometer and a set of cryostats/cryofurnaces covering a temperature range of 2 to 800 K. The measurements were performed at a photon energy of 12.398~keV (corresponding to a wavelength of 1~\AA) in vertical scattering four-circle geometry. In this configuration the optimum longitudinal resolution is $\delta Q \approx 3 \times 10^{-3}$~\AA $^{-1}$. Higher order x-rays were suppressed by the beamline mirrors and by the discriminator window setting. The samples were mounted on the tip of Cu sample holders for measurements in transmission (Laue) geometry when studying basal plane $(hk0)$ Bragg reflections. Naturally, $c$-axis reflections of type $(00l)$ could only be measured in reflection geometry. However, the thickness of the samples roughly corresponded to the x-ray penetration depth, ensuring that in both scattering geometries the bulk of the crystal was investigated. The beam footprint on the sample was about 1~mm$^2$.

\begin{figure}[t]
\begin{center} 
\includegraphics[scale=1]{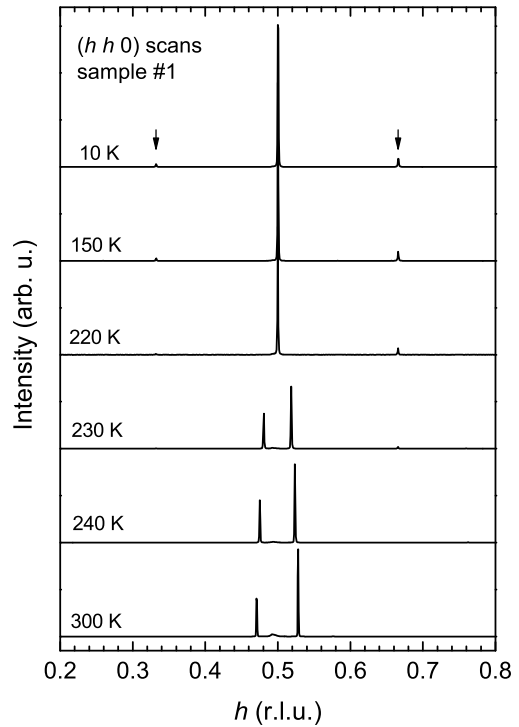} 
\caption{Selected longitudinal scans along $(h h 0)$, with $h = 0.2 - 0.8$ for sample~\#1 at various temperatures. The arrows mark reflections at $h \approx 1/3$ and $2/3$. A small, broad feature at $h = 0.495$ is associated with the (220) reflection of a highly textured Co$_3$O$_4$ contamination.}
\label{Fig1}
\end{center}
\end{figure}

Sample quality was checked at room temperature by measuring the width of the $(00l)$ reflections. No splitting of these reflections was observed in longitudinal scans and their width was resolution limited, showing that the $x$ value was homogeneous over the full volume. Rocking scans resulted in a mosaic spread of about 1$^\circ$ for $(00l)$ and 0.4$^\circ$ for $(hk0)$ reflections. Prior to synchrotron x-ray diffraction experiments, crystal \#1 was characterized using a conventional laboratory four-circle x-ray diffraction instrument equipped with a Mo source (Mo-K$_\alpha$ = 17.4 keV). The room-temperature lattice parameters $a = 2.814(8)$~\AA, $c = 11.13(3)$~\AA~ were determined on the basis of the basic hexagonal structure of Na$_x$CoO$_2$, in close agreement with literature data for Na$_{0.5}$CoO$_2$.\cite{Huang:2004} 
Using the established relation\cite{Foo:2004,Shu:2007} between the $c$ axis parameter and $x$, from the more accurate room temperature $c$-values measured in the synchrotron experiment, namely $c = 11.097(5)$\AA~ and $c = 11.105(5)$\AA, we arrive at $x = 0.530(5)$ and $x = 0.525(5)$ for samples \#1 and \#2, respectively. We associate the discrepancies between the $x$ values derived from NAA and the $c$ axis parameters with the presence of sizable amounts of secondary cobalt oxide, which is actually detected in the present diffraction experiments (see below). Since NAA determined the overall Na:Co ratio of the samples, Co in secondary phases artificially reduced the NAA $x$ values.

For comparison with the superstructures reported for other $x$ values, we will mainly use a notation related to the basic hexagonal cell $[a, c]$ in the following for indexing Bragg reflections $(h k l)$. The corresponding orthorhombic unit cell $[a^\prime, b^\prime, c^\prime]$ of the $x= 0.5$ phase is related to the basic hexagonal cell by $a^\prime = \sqrt3 a, b^\prime = 2a, c^\prime = c, \alpha = \beta = \gamma = 90^\circ, b^\prime \parallel a$. In this notation, for example, the reflection $(110)$ corresponds to $(040)^\prime$. Note that due to 120$^\circ$ twinning, different Bragg reflections of the orthorhombic phase may overlap in the bulk crystal, such as $(040)^\prime$ and $(320)^\prime$.

\section{Results}
\subsection{sample~\#1}

In a first round of experiments, the original sample~\#1 was studied in the temperature range 300~K~-~550~K. 
In a survey of reciprocal space carried out at 300~K we observed a number of strong superstructure reflections like $(\frac{1}{2} 0 0)$, $(\frac{1}{2} \frac{1}{2} 0)$, and $(\bar{\frac{1}{4}} \frac{3}{4} 0)$ [corresponding to	
$(1 0 0)^\prime$, $(0 2 0)^\prime$, and $(1 1 0)^\prime$], confirming that the crystal structure for the original sample~\#1 was the orthorhombic superstructure reported previously for Na$_{0.5}$CoO$_2$.\cite{Huang:2004} 
The strongest of these reflections, $(0 2 0)^\prime$, had about 6\% of the intensity of the related $(0 4 0)^\prime$. The reflections $(\frac{1}{2} 0 0)$ and $(\bar{\frac{1}{4}} \frac{3}{4} 0)$ were measured on heating and were found to vanish around 535~K, marking a transition to a disordered hexagonal phase. This transition was reversible. However, during several heating-cooling cycles around 500~K, we noted a continuous reduction of the intensities of these reflections and finally they were drastically reduced after cooling the sample back to 300~K.

\begin{figure}[t]
\begin{center} 
\includegraphics[scale=0.9]{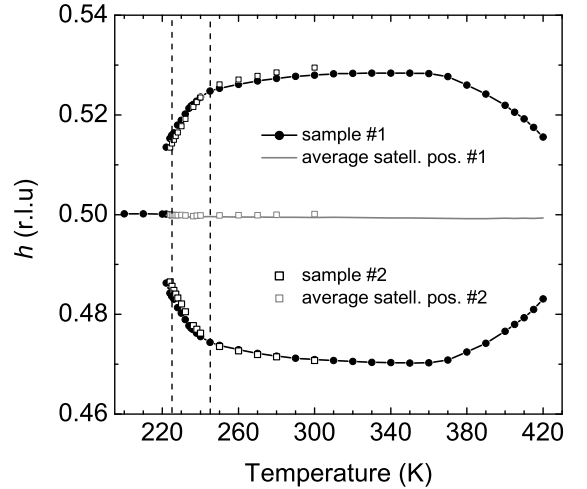} 
\caption{Temperature dependence of the positions of the central $(\frac{1}{2} \frac{1}{2} 0)$ and the incommensurate satellites $(h h 0)$ for sample~\#1 (solid symbols). The maximum splitting corresponds to a propagation vector $q = 0.529$. The grey line indicates the average between the two satellite positions, reflecting the lattice expansion, which actually is negligible. Open symbols show the related data for sample~\#2. Dashed vertical lines indicate characteristic temperatures discussed for sample~\#2.}
\label{Fig2}
\end{center}
\end{figure}

A subsequent $h$-scan along the $(h h 0)$ direction at 300~K showed the presence of a pair of satellite reflections at $h= 0.475$ and 0.525, while the central reflection $(\frac{1}{2} \frac{1}{2} 0)$ was absent (see FIG.~\ref{Fig1}). Obviously, excessive heating lead to a modification of the room temperature Na ordering of sample~\#1 from the orthorhombic $x = 0.5$ phase to an incommensurate phase.\cite{footnote} Sample~\#2 exhibited the same incommensurate superstructure, similar to those reported before,\cite{Zandbergen:2004,Yang:2005} without any additional heat treatment (see below). The heat treatment of sample~\#1 did not result in any significant change of the $c$ axis parameter to within 0.01~\AA, hence the variation of the Na concentration was $\leq 0.01$.

Scans along $(h h 0)$, with $h = 0.2 - 0.8$, were measured on sample~\#1 at various temperatures (see FIG.~\ref{Fig1}). 
We observe a phase transition around 225~K. Below 225~K we find Bragg reflections close to the commensurate positions $h = \frac{1}{3}, \frac{1}{2}$, and $\frac{2}{3}$, while above that temperature, only the pair of incommensurate satellites mentioned above is visible (besides a contamination from Co$_3$O$_4$). This observation immediately suggests a phase-segregation below 225~K into two different phases, which we will label A and B in the following,  producing 1/2- and 1/3-integer type of reflections $(h h 0)$, respectively. 

The temperature dependence of the splitting of the incommensurate satellites $(h h 0)$ was followed on sample~\#1 between 200~K and 450~K (FIG.~\ref{Fig2}). The maximum splitting is observed around 350~K, corresponding to a propagation vector $q = 0.529$. Notably, the incommensurate ordering vanishes above 420~K, roughly consistent with the previous electron diffraction work.\cite{Yang:2005} In the course of these measurements we found indications of a continuous sample degradation when heating the sample, like an increase of the amount of Co$_3$O$_4$ contamination, a decrease of the intensities of the incommensurate satellites and also a slight increase of the maximum peak splitting. Finally, after several cycles around 450~K, the incommensurate phase was destroyed.

\begin{figure}
\includegraphics[scale=1]{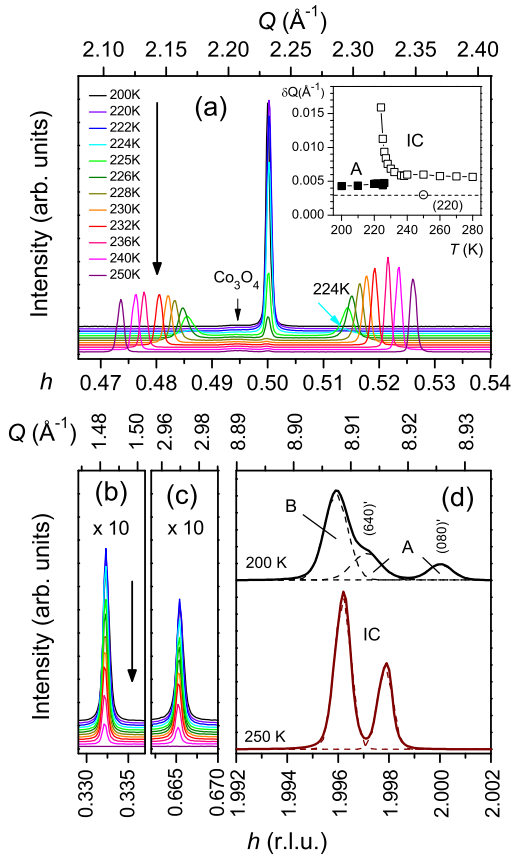}
\caption{(Color online) Selected longitudinal scans along $(h h 0)$ for sample~\#2 at various temperatures. Absolute $Q$ scales are given at the top of each panel. The $h$ indexing (bottom scales) is normalised to the position of the $(\frac{1}{2} \frac{1}{2} 0)$ reflection at 200~K. Panel (a) shows the $(\frac{1}{2} \frac{1}{2} 0)$ from the orthorhombic phase~A that splits into two satellites around 225~K with a co-existence region between 224~K and 228~K. The inset shows the FWHM linewidths for both types of reflections (exp. resolution 0.003~\AA $^{-1}$). Panels (b) and (c) show the reflections from the second low-$T$ phase~B. These reflections vanish only above 245~K. The deviation from the ideal 1/3 and 2/3 positions is related to a larger basal plane lattice parameter. Panel (d) shows the (220) region at 200~K and 250~K. At 200~K a superposition of the (640)' and (080)' reflections from phase~A (intensity ratio 2:1) and the (220) from the hexagonal phase~B is observed. At 250~K, where only the IC phase is present, a 2:1 doublet is found, pointing to an orthorhombic distortion. Different data curves are offset for clarity.}
\label{Fig3}
\end{figure}

\subsection{sample~\#2}

The transition between the phase segregated and the incommensurate states was studied in more detail on the as-prepared sample~\#2. To avoid sample damage, the temperature range was limited to $T \leq 295$~K in these measurements. FIG.~\ref{Fig3} shows the temperature evolution of Bragg reflections of type $(h h 0)$ for the various phases between 200~K and 250~K. On heating, the central $(\frac{1}{2} \frac{1}{2} 0)$ splits into two satellites at 224~K, marking the commensurate-to-incommensurate transition. Between 224~K and 228~K the central $(\frac{1}{2} \frac{1}{2} 0)$ co-exists with the incommensurate satellites. Above 228~K only a tiny residual of the central reflection remains visible, while the splitting of the pair of incommensurate satellites increases continuously. The temperature dependence of the satellite positions is very similar to the one observed for sample~\#1 (FIG.~\ref{Fig2}). The hysteresis of the commensurate-to-incommensurate transition was measured by monitoring the intensity at $(\frac{1}{2} \frac{1}{2} 0)$ while cycling the temperature down and up at a rate of 5~K/min between 200 and 230~K (not shown). The data reveal a hysteresis of width 5~K, thus the commensurate-to-incommensurate transition is of first order type.

We note in passing that in the phase segregated state we also observe $(h00)$ type Bragg reflections like $(\frac{1}{2} 0 0)$ and $(\frac{1}{3} 0 0)$ which vanish on heating around 225~K and 245~K, respectively (not shown), consistent with the behavior of the related $(hh0)$ reflections. Above 245~K, no $(h00)$ type reflection is observed for the full range $0 < h < 1$.

From the absolute $Q$ value of the central $(\frac{1}{2} \frac{1}{2} 0)$ reflection we infer a value of $b^\prime = 5.632(1)$~\AA\ at 200~K, corresponding to a quasi-hexagonal lattice parameter $a = 2.8160(5)$~\AA\ of phase~A. Assuming that the reflections from the second low-$T$ phase are related to a commensurate value 1/3, we infer a somewhat larger value $a = 2.823(1)$~\AA, which is a first indication that phase~B has a Na concentration $x > 1/2$.

\begin{figure}[t] 

\includegraphics[scale=1]{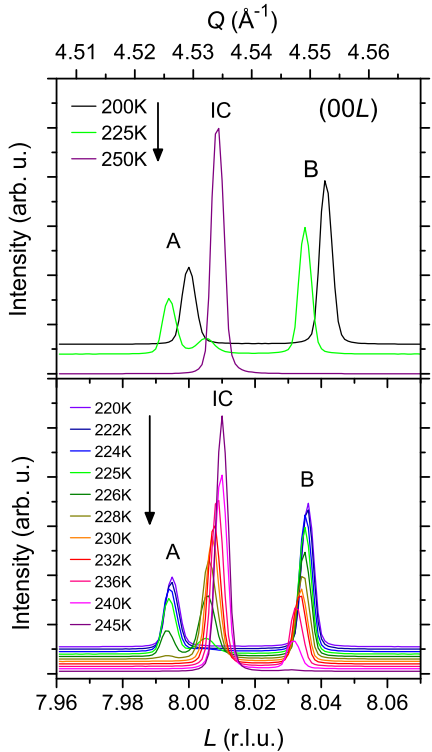}
\caption{(Color online) Selected longitudinal scans along $(0 0 l)$ for sample~\#2 at various temperatures. The absolute $Q$ scale is given at the top of the figure. The $l$ indexing (bottom scale) is normalised to the position of the (008) of phase~A at 200~K. Top panel: Three $l$-scans, (i) well below (200~K), (ii) right at (225~K) and (iii) well above (250~K) the commensurate-incommensurate transition. At 200~K the reflection is split into two components, which are assigned to the two phases A and B, respectively, see text. At 225~K all three phases coexist. At 250~K only a single reflection from the IC phase is observed. The $T$-dependent shift of the commensurate lines is caused by thermal expansion. Bottom panel: Same kind of data, but in narrow $T$ steps around the transition. The different data curves are offset for clarity.}
\label{Fig4}

\end{figure}

\begin{figure}[t]
\includegraphics[scale=1]{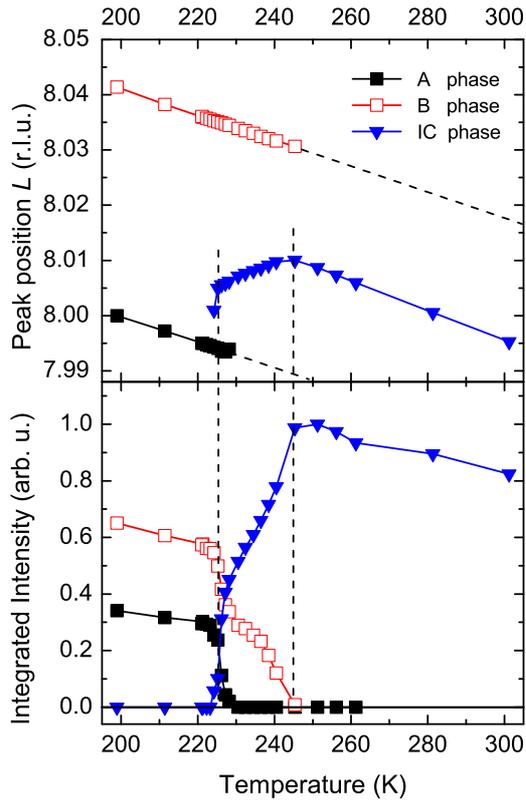}
\caption{(Color online) Temperature dependence of (top) the positions and (bottom) the integrated intensities of the (008) type Bragg reflections for the various phases. The peak positions are normalised to the (008) of phase~A at 200~K. The shift of the commensurate peak positions to lower $L$ values on increasing $T$ is caused by thermal expansion. The dashed lines show the extrapolation of thermal expansion to 300~K.}
\label{Fig5}
\end{figure}

A more accurate information on the Na stoichiometry of phase~B and the IC phase can be obtained by comparing the $c$-axis parameters of the various phases. This was done by studying the temperature evolution of the (008) reflection across the commensurate-to-incommensurate transition (FIG.~\ref{Fig4} and FIG.~\ref{Fig5}). Above 250~K only a single reflection is observed, which shows that the incommensurate phase is homogeneous in this temperature range. Actually, we have checked a much broader region $l = 7.7 - 8.6$ (not shown), with no indication of another reflection.
In contrast, at 200~K, well below the transition, the (008) reflection is split into two components. The peak with the smaller $Q$ value vanishes above 230~K, while the peak at larger $Q$ only vanishes above 245~K. Comparing this behavior with that of the related $(h h 0)$ reflections (FIG.~\ref{Fig3}), we associate the first peak with the orthorhombic phase~A and the latter with phase~B. The width of these reflections is resolution limited, thus for all three phases the structural coherence length along the $c$ direction is $> 5000$~\AA.

At 200~K, the absolute $c$ axis parameters are $c = 11.098$~\AA\ for the phase~A and $c = 11.042$~\AA~for phase~B. From the shift of the (008) reflections between 200~K and 225~K (see top panel of FIG.~\ref{Fig5}) we derive a thermal expansion coefficient for the $c$ axis of ${\rm d}c/c{\rm d}T = 2.86\times 10^{-5}$~K$^{-1}$. Assuming a $T$-linear thermal expansion between 200~K and 300~K, the \textit{extrapolated} room temperature $c$ axis parameters are 
$c = 11.129$~\AA\ for phase~A and $c = 11.073$~\AA~for phase~B. The former value is in perfect agreement with the literature data for Na$_{0.5}$CoO$_2$.\cite{Huang:2004} Using again the established linear relation\cite{Lang:2008} between the $c$ axis parameter and $x$, we arrive at a Na concentration $x = 0.55$ for phase~B.

The temperature dependence of the integrated intensities of the (008) type Bragg reflections for the various phases (bottom panel of FIG.~\ref{Fig5}) can be interpreted as the evolution of the relative volume fractions, considering that the integrated intensities of the corresponding reflections are a direct measure of the associated sample volume. Taking into account a small increase of the (008) structure factor by 5\% when increasing $x$ from 0.50 to 0.55, we get a value of 1:1.8 for the ratio of the volumes of phases A and B at 200~K. Using these values, sodium balance would result in a value $x = 0.532$ for the average sodium concentration of sample~\#2. On heating, the onset of incommensurate ordering around 225~K is associated with a collapse of phase~A. The volume fraction of phase~B shows a sudden drop by 35\% at that transition. In the temperature region between 230~K and 245~K, the volume fraction of the IC phase is continuously increasing at the cost of phase~B until the latter vanishes above 245~K. Interestingly, in this $T$ region the (008) reflection from the IC phase is shifting to larger $Q$ values (see top panel of FIG.~\ref{Fig5}), opposite to the trend from thermal expansion observed above 245~K. Assuming a linear relation between $c$ and $x$,\cite{Lang:2008} the behavior of the (008) peak positions suggests that above 245~K the Na concentration of the IC phase is the 1:1 average of the Na concentrations of phases A and B, i.e. $x = 0.525$, slightly smaller than expected fron the sodium balance. It appears that the Na balance is not strictly conserved, i.e., some sodium hides in grain boundaries or in a yet undetected phase.

Finally, we discuss the data for the (220) region (FIG.~\ref{Fig3}(d)). Due to twinning, we expect a splitting of the quasi-hexagonal (220) reflection into a pair of reflections (640)' and (080)' for the orthorhombic phase~A. We assign the pair of lines at $h = 1.997$ and 2.000 with these reflections, since they have the expected 2:1 intensity ratio. From the absolute $Q$ values we derive lattice parameters $a^\prime = 4.887(1)$~\AA\ and $b^\prime = 5.632(1)$~\AA\ at 200~K for phase~A. Thus, for the orthorhombic distortion we obtain $2 a^\prime /\sqrt 3 b^\prime = 1.002$, i.e., a slight compression along the $b^\prime$ axis. The single reflection at 1.996 is associated with phase~B, which apparently is hexagonal, with $a = 2.823(1)$~\AA. The relative peak intensities are consistent with the 1:1.8 ratio of the volumes of phases A and B determined above. At 250~K we observe a pair of reflections with a 2:1 intensity ratio. Since at that temperature only the IC phase is present, we conclude that this phase also is orthorhombic. From the absolute $Q$ values we derive lattice parameters $a^\prime = 4.888(1)$~\AA\ and $b^\prime = 5.638(1)$~\AA\ at 250~K for the IC phase. Interestingly, $a^\prime$ is practically the same as for phase~A, but the orthorhombic splitting is only about half as big. Note that the lattice expansion of the basal plane is negligible (see FIG.~\ref{Fig2}), thus the given values may be extrapolated to room temperature.

It is also interesting to compare the linewidth of the various $(hh0)$ reflections. At 250~K, the (640)'/(080)' pair of the IC phase (FIG.~\ref{Fig3}(d)) exhibits a FWHM of $\delta Q = 3 \times 10^{-3}$~\AA $^{-1}$, while at 200~K the FWHM of each reflection is $\delta Q^\prime = 4.5 \times 10^{-3}$~\AA $^{-1}$. This points to a reduced structural basal plane coherence length in the phase segregated state. In the IC phase the reflections are resolution limited, thus from the line broadening we derive an upper limit of $2\pi/\sqrt{\delta Q^{\prime~2} -\delta Q^2} = 1900$~\AA\ for the average basal plane crystal domain size at 200~K. The linewidth of the IC $(hh0)$ satellites yields information on the coherence length of the structural modulation along the $b^\prime$ axis of the IC phase (FIG.~\ref{Fig3}(a), inset). Close to the transition, this coherence length is strongly reduced, starting at a value of about 400~\AA\ at 225~K, while above 240~K it levels off at a value of 1200~\AA.

\section{Discussion}

There is no doubt that phase~A is the well-known orthorhombic $x = 0.50$ phase reported previously.\cite{Huang:2004} To our knowledge, we obtained the first quantitative determination of the orthorhombic distortion of this phase, $2 a^\prime /\sqrt 3 b^\prime = 1.002$. 

\begin{figure}[t]
\includegraphics[scale=0.8]{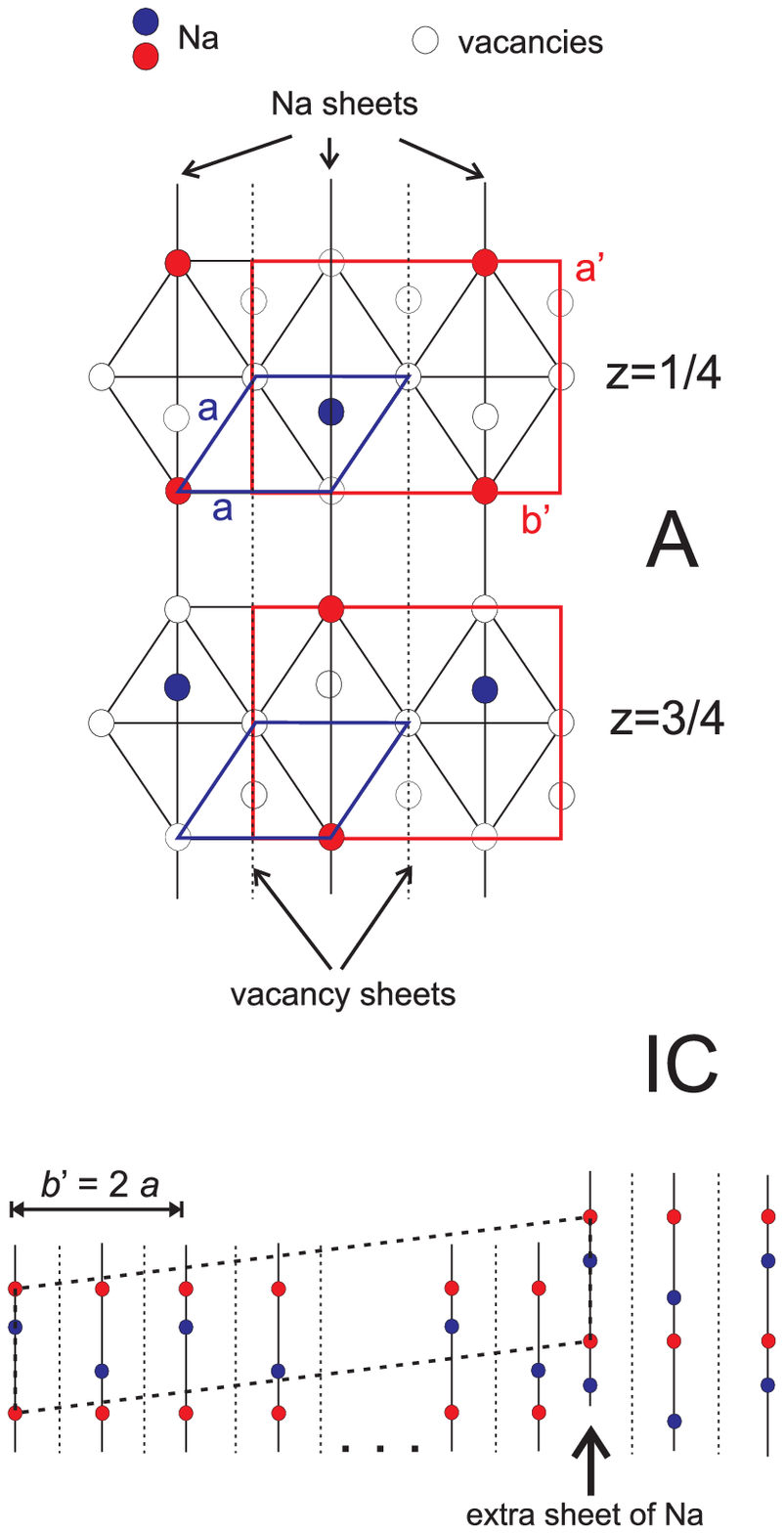}
\caption{(Color online) Structure models for phase~A with $x = 0.50$ (top, after Huang \textit {et al.}\cite{Huang:2004}) and the IC phase with $x = 0.5 + \delta$ (bottom). The latter model, suggested previously,\cite{Lang:2008} is very similar to that of Zandbergen {et al.},\cite{Zandbergen:2004} which assumes extra sheets of vacancies to account for sodium concentrations $x = 0.5 - \delta$.}
\label{Fig6}
\end{figure}

Concerning the IC phase it is important to note the close agreement of the room temperature values of $x = 0.525$ and $q = 0.529$ for sample \#2. This is no accident but leads to a simple model for the structure of the IC phase: In the original electron diffraction work a model for the incommensurate modulation has already been suggested assuming $x < 0.5$.\cite{Zandbergen:2004} This model involves extra sheets of vacancies (in planes spaned by the [100] and [001] axes) that are distributed regularly along the modulation direction [110]. This model naturally results in the coincidence $x = q$. One can adopt this model to the actual case of $x > 0.5$ by replacing the "extra sheets of vacancies" by "extra sheets of Na ions" (see FIG.~\ref{Fig6}), as has been suggested previously.\cite{Lang:2008}

The question of the structure of phase~B is more difficult to answer. For this phase we have found a sodium concentration $x = 0.55$. Quite a number of theoretical studies dealt with the prediction of possible sodium ordered ground states for the full range of concentrations $x = 0.11$ - 1.\cite{Wang:2007, Meng:2008, Wang:2008} Experimental studies mainly focused on the range $x \geq 2/3$, where many stable phases at rational Na occupancies $x$ have been reported at ambient temperature.\cite {Shu:2007,Chou:2008,Huang:2009,Shu:2009,Platova:2009}
However, only a few studies deal with the $x$-range of interest. Theoretically, for $x = 5/9$ a possible superstructure has been predicted\cite{Meng:2008} that would result in a set of Bragg reflections $(h h 0)$ with $h = n/9$. A strange behavior was observed experimentally in a sample with $x = 0.58$. Here, a $\sqrt{7} a \times \sqrt{7} a$ superstructure, corresponding to $x = 4/7$, exists only for a limited temperature range 235~K~$< T <$~288~K.\cite{Igarashi:2008} A Raman spectroscopy study on a sample with $x = 0.56$, in which a Na rearrangement transition around 240~K is found, has been interpreted as a transition from an orthorhombic low-$T$ to a hexagonal high-$T$ structure.\cite{Wu:2008} All this is in contrast to our findings, thus none of these reports applies to the present results.

The only relevant information again stems from the original electron diffraction results on a sample exhibiting the IC modulation.\cite{Zandbergen:2004} As mentioned above, at 100~K this sample contained domains showing an apparent tripling of the \textit{orthorhombic} unit cell along the $a^\prime$ and $b^\prime$ axes. This sample probably  exhibited a similar phase-segregation as reported in the present work, although $x < 0.5$ was assumed and the set of observed ($h h 0$) reflections differs (e.g., $(\frac{1}{4} \frac{1}{4} 0)$ is strong in the previous but absent in the present work). 

Motivated by the electron diffraction work we have carried out a more thorough search for superstructure reflections from phase~B at 200~K. Besides the dominant $1/3$-integer type basal plane reflections ($hk0$), we also find $1/6$-integer reflections, like the very weak $(\frac{1}{6} \frac{1}{6} 0)$, stronger $(\frac{1}{6} \frac{1}{3} 0)$, $(\frac{5}{6} \frac{1}{3} 0)$ and quite significant $(\frac{1}{6} \frac{1}{2} 0)$, $(\frac{2}{3} \frac{5}{6} 0)$ reflections. Although the detailed determination of the structure of phase~B is beyond the scope of the present work, on the basis of the present (limited) data we conlude that phase~B appears to be hexagonal and most probably involves a $6a \times 6a \times c$ unit cell. This cell could accomodate a stoichiometry $x = 5/9$. Thus we suggest that phase~B observed in the present work is a new Na ordered phase with $x = 5/9$ that is only stable below 245~K. Remarkably, a recent theoretical study, in which a temperature-concentration phase diagram for Na$_x$CoO$_2$ was suggested, predicted a stable $x = 5/9$ phase below 260~K, in very good agremment with our result.\cite{Hinuma:2008} This work also indicated that the above mentioned theoretical superstructure~\cite{Meng:2008} was not the true ground-state configuration.

The most interesting feature of the present work is the commensurate-to-incommensurate transition, whose driving force is most probably entropy. 
Starting from the phase segregated state, around 225~K phase~A collapses and IC ordering sets in. This is remarkable, because previously the stoichiometry $x = 0.5$ has been suggested to be very stable. In addition, no theoretical work has predicted any type of incommensurate Na ordering.

We interprete the observed temperature dependence of the (008) positions and intensities (Fig.~\ref{Fig5}) with the following scenario: In a first step, around 225~K Na ions from a 35\% volume fraction of grains from phase~B travel into neighboring grains of phase~A to form the new IC phase in the sum of their volumes. In a second step, between 230~K and 245~K, Na ions travel from phase~B into neighboring grains of the new IC phase to equilibrate the Na concentration. In this $T$ region, not only the volume fraction but also the Na concentration in the IC phase is steadily inreasing, resulting in a contraction of the $c$ axis parameter. Finally, at 245~K there is no phase~B left and the full sample volume is in a homogeneous IC state. The observed behavior requires a very high long-range sodium mobility above 200~K, not only within the hexagonal basal plane but also along the $c$ direction. This is consistent with previous conjectures.~\cite{Weller:2009, Julien:2008, Lang:2008}

It is interesting to note that incommensurate Na ordering at room temperature has been reported for various ranges of sodium concentrations from x-ray powder diffraction.\cite{Lang:2008, Platova:2009} However, in the $x$ range of interest, the IC state appears to be stable only for a very narrow range $0.52 \leq x \leq 0.53$, while a neighboring range, $0.58 \leq x \leq 0.62$, is significantly broader. This may explain why initially sample \#1 did not show the IC structure although the Na concentration was very close to that of sample \#2 and why heat treatment may destroy the IC phase. In the powder diffraction work\cite{Lang:2008} a sample with $x = 0.55$ did not exhibit any sodium ordering at room temperature. Interestingly, this sample exhibited an anomaly of the susceptibility around 230~K and NMR revealed three-dimensional correlations between Na ions at low $T$. Based on the present results we suspect that also this sample exhibited sodium ordering at low temperatures. 

\section{Conclusions}

We have shown that at room temperature Na$_{x}$CoO$_2$ with composition $x \approx 0.53$ exhibits homogeneous incommensurate Na ordering with $q \approx x$. On cooling, unusual reversible phase-segregation into two volume fractions takes place, where one volume fraction shows the well-known $x = 0.50$ superstructure, while a second volume fraction with $x = 0.55$ exhibits commensurate sodium ordering, presumably with a $6a \times 6a \times c$ hexagonal supercell. The commensurate-to-incommensurate transition is an intrinsic feature of samples with Na concentrations $x = 0.5 + \delta$ with $\delta \approx 0.03$, but also for other sodium concentrations the possibility of temperature dependent phase-segregation effects should be taken into account.

\hspace{0.1cm}

\section{Acknowledgements}

A.U.B.W. was supported by the DFG under Contract No. SU229/8-1. Construction of the beamline MAGS has been founded by the BMBF via the HGF-Vernetzungsfonds under contracts No.~01SF0005 and 01SF0006. We acknowledge stimulating  discussions with D.J.P. Morris.


\begin{thebibliography}{50}
\bibitem{footnote} We reproduced the transformation from the orthorhombic to the incommensurate room temperature structure by heat treatment above 500~K on a second piece of batch \#1 in an extensive laboratory x-ray diffraction experiment.
\end{thebibliography}
\end{document}